\documentclass[11pt,english,aps,prc,showpacs,letter,longbibliography]{revtex4-1}
\usepackage{amsmath}
\usepackage[    bookmarks,
                 bookmarksopen = true,
                 bookmarksnumbered = true,
                 linktocpage,
                 colorlinks = true,
                 linkcolor = blue,
                 urlcolor  = blue,
                 citecolor = blue,
                 anchorcolor = green,
                 hyperindex = true,
                 hyperfigures]
        {hyperref}

\usepackage{enumitem}
\usepackage{graphicx}
\usepackage{esint}
\usepackage[utf8]{inputenc}
\usepackage[T1]{fontenc}
\usepackage{float}
\usepackage{fancyhdr}

\usepackage{ulem}
\usepackage{lineno}

\usepackage{listings}

\DeclareFixedFont{\ttb}{T1}{txtt}{bx}{n}{8} 
\DeclareFixedFont{\ttm}{T1}{txtt}{m}{n}{8}  

\usepackage{color}
\definecolor{deepblue}{rgb}{0,0,0.5}
\definecolor{deepred}{rgb}{0.6,0,0}
\definecolor{deepgreen}{rgb}{0,0.5,0}

\newcommand\pythonstyle{\lstset{
language=Python,
basicstyle=\ttm,
otherkeywords={self},             
keywordstyle=\ttb\color{deepblue},
emph={MyClass,__init__},          
emphstyle=\ttb\color{deepred},    
stringstyle=\color{deepgreen},
frame=tb,                         
showstringspaces=false            %
}}

\lstnewenvironment{python}[1][]
{
\pythonstyle
\lstset{#1}
}
{}

\newcommand{\ba}{\begin{eqnarray}}
\newcommand{\ea}{\end{eqnarray}}

\makeatletter
\usepackage{soul}

\makeatother

\begin{document}

\title{Deep learning assisted jet tomography for the study of Mach cones in QGP}

\author{Zhong Yang$^1$, Yayun He$^{2,3}$, Wei Chen$^{4}$,  Wei-Yao Ke$^{5,6,7}$, Long-Gang Pang$^{1}$\footnote{email: lgpang@ccnu.edu.cn} and Xin-Nian Wang$^{1,5,6}$\footnote{email: xnwang@.lbl.gov, Current address$^6$.}
}

\address{$^{1}$Key Laboratory of Quark \& Lepton Physics (MOE) and Institute of Particle Physics, Central China Normal University, Wuhan 430079, China}

\address{$^2$Guangdong Provincial Key Laboratory of Nuclear Science, Institute of Quantum Matter, South China Normal University, Guangzhou 510006, China}

\address{$^3$Guangdong-Hong Kong Joint Laboratory of Quantum Matter, Southern Nuclear Science Computing Center, South China Normal University, Guangzhou 510006, China}

\address{$^4$School of Nuclear Science and Technology, University of Chinese Academy of Sciences, Beijing 100049, China}

\address{$^{5}$Physics Department, University of California, Berkeley, CA 94720, USA}

\address{$^{6}$Nuclear Science Division, Lawrence Berkeley National Laboratory, Berkeley, CA 94720, USA}

\address{$^{7}$Theoretical Division, Los Alamos National Laboratory, Los Alamos, NM 87545, USA}

\begin{abstract}

Mach cones are expected to form in the expanding quark-gluon plasma (QGP) when energetic quarks and gluons (called jets) traverse the hot medium at a velocity faster than the speed of sound in high-energy heavy-ion collisions. The shape of the Mach cone and the associated diffusion wake are sensitive to the initial jet production location and the jet propagation direction relative to the radial flow because of the distortion by the collective expansion of the QGP and large density gradient.
The shape of jet-induced Mach cones and their distortions in heavy-ion collisions provide a unique and direct probe of the dynamical evolution and the equation of state of QGP. However, it is difficult to identify the Mach cone and the diffusion wake in current experimental measurements of final hadron distributions because they are averaged over all possible initial jet production locations and propagation directions. To overcome this difficulty, we develop a deep learning assisted jet tomography which uses the full information of the final hadrons from jets to localize the initial jet production positions. This method can help to constrain the initial regions of jet production in heavy-ion collisions and enable a differential study of Mach-cones with different jet path length and orientation relative to the radial flow of the QGP in heavy-ion collisions. 
\end{abstract}

\keywords{Heavy ion collisions, Deep learning, Mach cones, jet energy loss}

\pacs{}

\maketitle

\section{Introduction}

All projectiles moving through medium feel resistance and eventually slow down, depending on the density of the medium and the frictional force.  Typical bullets, for example, lose all their energy to the friction and come to a stop after about only a few feet of distance in the water which is about 800 times denser than the air.  The lost energy generates sound waves in the medium that travels with the projectiles and leaves behind a diffusion wake. When the speed of a fast-moving projectile exceeds the velocity of sound in the medium, fronts of the sound waves are squeezed  to form a Mach cone or the sonic boom. The same phenomenon is also expected in a small droplet of subatomic liquid \cite{Baumgardt:1975qv,Rischke:1990jy,Stoecker:2004qu,Casalderrey-Solana:2004fdk,Satarov:2005mv,Ruppert:2005uz, Gubser:2007ga, Qin:2009uh,Neufeld:2009ep,Bouras:2012mh,Yan:2017rku,Casalderrey-Solana:2020rsj,JETSCAPE:2020uew, Casalderrey-Solana:2020rsj,Stoecker:2004qu} called quark gluon plasma (QGP) the size of an atomic nucleus in high energy nuclear collisions. In contrast to the normal nuclear matter where quarks and gluons are confined to the domains of each nucleon (proton or neutron), they can roam freely in the whole volume of QGP. In a process called jet quenching \cite{Bjorken:1982tu,Gyulassy:1990ye,Wang:1992qdg,Qin:2015srf}, energetic quarks and gluons (also called partons) or jets are produced through hard scattering of beam partons from the colliding nuclei before the formation of QGP. These fast jets have to traverse the QGP at the speed of light ($c$), which is faster than the upper bound of the velocity of sound ($c_s < {c/\sqrt{3}}$ ) in QGP \cite{HotQCD:2014kol}. As the fast jets lose energy-momentum along their trajectories through jet-medium interaction, their lost energy will induce supersonic sound waves in the form of Mach cones on a femtometer scale inside the QGP.

\begin{figure}
    \centering
    \includegraphics[width=0.6\textwidth]{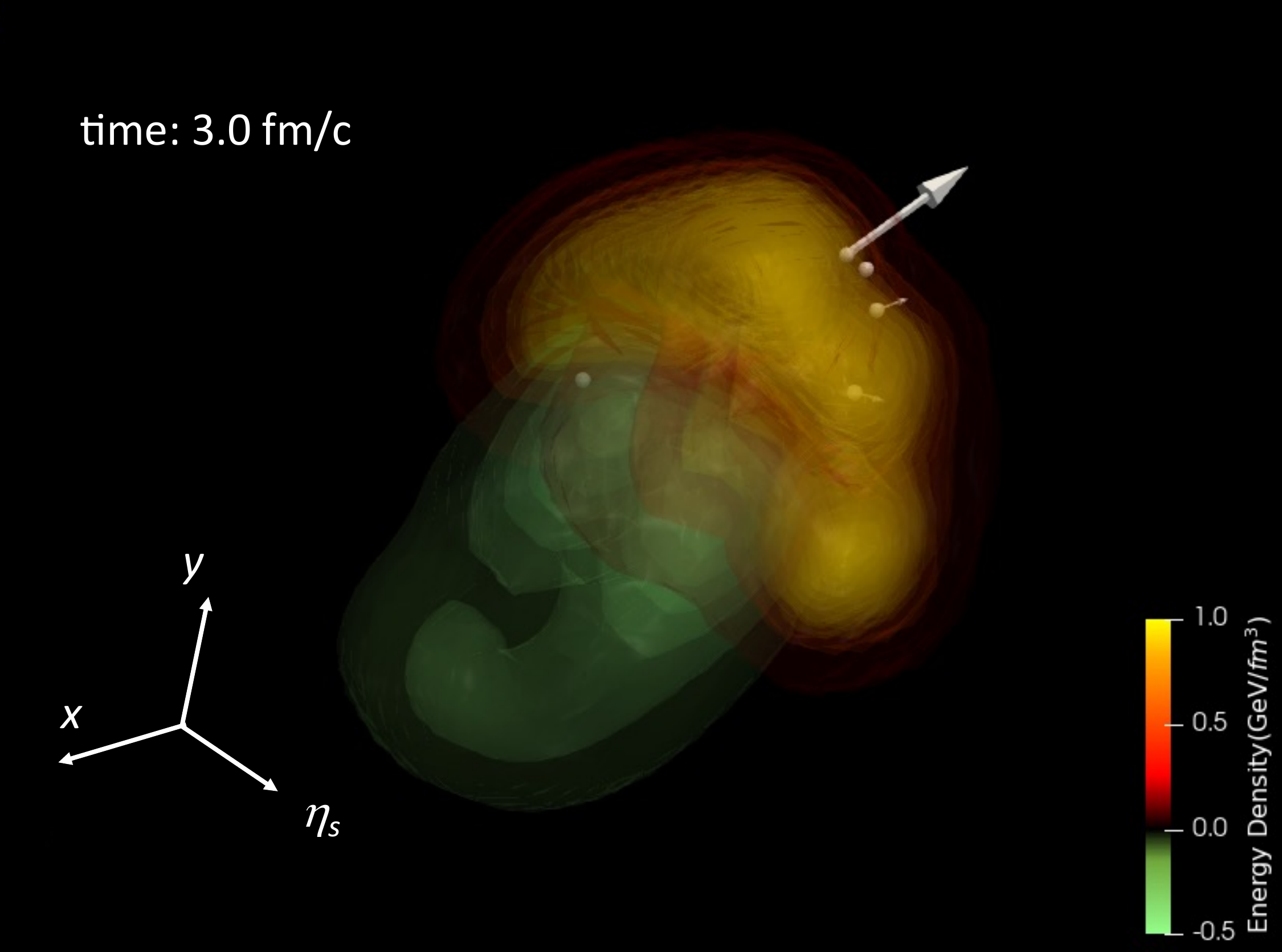}
    \caption{(color online) A snapshot of the 3-dimensional energy density distribution in the transverse plane $(x, y)$ and the spatial rapidity ($\eta_s$) of the medium response induced by a $\gamma$-triggered jet at time $\tau=3$ fm/$c$ in a central Pb+Pb collisions at $\sqrt{s}=5.02$ TeV simulated by the CoLBT-hydro model (see description in Sec. II). White dots represent the position of jet shower partons and the arrows their momenta.}
    \label{fig:3dmachcone}
\end{figure}

The jet-induced subatomic Mach cones can be studied theoretically using transport models \cite{Ma:2006fm,Li:2010ts,Ma:2010dv,Luo:2021voy,Cao:2020wlm}, linearized hydrodynamics \cite{Casalderrey-Solana:2004fdk,Neufeld:2008dx,Casalderrey-Solana:2020rsj,Ruppert:2005uz,Gubser:2007ga,Qin:2009uh,Ayala:2016pvm,Casalderrey-Solana:2020rsj,Cao:2020wlm,Neufeld:2008fi,Yan:2017rku} 
as well as full hydrodynamics \cite{Betz:2008ka,Tachibana:2015qxa,Chen:2017zte,Hirano:2017wsq,JETSCAPE:2020uew,Cao:2020wlm,Du:2022oaw,Karpenko:2019xsc,Chaudhuri:2006qk,Chaudhuri:2007vc,Du:2022oaw} as illustrated in Fig.~\ref{fig:3dmachcone}. These Mach cones are different from those in a classical and macroscopic medium. First of all, partonic jets are quantum projectiles. They not only lose energy-momentum through elastic scattering with the medium but also through self and medium-induced gluon radiation with intricate quantum interference due to multiple scattering in medium \cite{Gyulassy:1993hr,Wang:1994fx,Baier:1996kr}. This unique energy loss mechanism will lead to a deformed Mach cone \cite{Qin:2009uh,Neufeld:2009ep} if all or part of the radiative energy loss is converted into medium excitation. Secondly, the extreme hot and dense QGP droplet expands very rapidly due to large pressure and density gradient during the jet propagation. The induced Mach cone excitation will also be deformed by the large density gradient and the radial flow of the medium \cite{Chaudhuri:2006qk,Li:2010ts,Ma:2010dv,Tachibana:2015qxa}. Experimental observation and measurements of the jet-induced Mach cone excitation therefore can help to study the dynamical evolution and intrinsic properties of the QGP medium such as the temperature and flow profiles, the opacity to jet partons, the equation of state and transport coefficient of the QGP \cite{Stoecker:1986ci,Bouras:2012mh,Stoecker:2004qu}.

Since both jets and the QGP with jet-induced medium excitation will become hadrons in the final state, one can search and study the jet-induced Mach cone in heavy-ion collision experiments through jet and hadron correlations. Jets in the final state are represented by the energetic leading or clusters of hadrons and hadronization of QGP leads to a large number of soft hadrons as recorded by experimental detectors. Because of the large background of soft hadrons from QGP that are not related to jets and jet-induced Mach cone, it is rather difficult to unambiguously identify the signal of the Mach cone in experimental data.  For example, fluctuating initial energy density distributions and collective expansion of QGP can lead to nontrivial two-hadron correlations in  azimuthal angle from anisotropic flows
\cite{Schenke:2010rr,Ma:2010dv,Takahashi:2009na,Alver:2010gr,ALICE:2011ab,Cao:2020wlm} that can be misidentified as a possible signal of Mach cone.  Such background can be eliminated in neutral boson ($Z^0$) or photon ($\gamma$)-jet events since the $\gamma/Z^0$ trigger does not interact with the QGP medium and therefore is not correlated with the underlying soft hadrons. However, the most recent experimental data on $\gamma/Z^0$-hadron correlations \cite{CMS:2021otx} have not shown any visible and unambiguous signal of jet-induced Mach cones without any additional event selections.

The absence of clear experimental evidence for jet-induced Mach cones in $\gamma/Z^0$-jet events could be caused by averaging event over the initial jet production positions and propagation directions. For a given initial jet production position and propagation direction,  simulations show that jet-induced Mach cones lead to some unique angular distributions of soft hadrons due to distortion by the density gradient and radial flow in an expanding QGP medium \cite{He:2015pra,Chen:2017zte,Yang:2021qtl}. However, for fixed transverse momentum of the trigger $\gamma/Z^0$ in experiments, the initial $\gamma/Z^0$-jet can be produced at any transverse position $(x,y)$ according to the distribution of the number of binary collisions in the overlap region of two colliding nuclei. The final $\gamma/Z^0$-hadron correlations measured in experiments are averaged over events with different $\gamma/Z^0$-jet production positions and jet propagation direction or effectively different jet propagation path length (different amount of jet energy loss), the temperature gradient, as well as the intersection between jet propagation and collective flow of the medium. The final averaged $\gamma/Z^0$-hadron correlation unfortunately loses many of the distinctive features from a typical Mach cone. 

Attempts have been made in the past to locate the initial jet production position using jet tomography.  Since $\gamma/Z^0$ does not lose energy in the QGP medium, the change of $\gamma/Z^0$-jet energy asymmetry $\chi = E_{\gamma/Z^0} - E_{\rm jet}$ is related to the jet energy loss which depends monotonically on the jet propagation length. It is therefore proposed in a longitudinal jet tomography \cite{Zhang:2009rn} that the $\gamma/Z^0$-jet energy asymmetry can be used to locate the initial longitudinal position (along the jet propagation direction) or the jet path length. Similarly, the transverse asymmetry $A_T$, which is defined to characterize the jet transverse diffusion due to the density transverse gradient of the medium, has been proposed in a transverse or gradient tomography  \cite{He:2020iow,Fu:2022idl} to approximately locate the jet initial transverse position perpendicular to the jet propagation.  The 2D jet tomography combining the longitudinal and transverse jet tomography is shown to be able to locate the initial jet production positions, including both the path length and perpendicular coordinates, and enhance detailed structures of the diffusion wake of the Mach cones in $\gamma$-hadron
correlations \cite{Yang:2021qtl,Yang:2022nei}. Recently, deep neural network was also employed to estimate the energy loss of hadron-triggered jet which can also tell roughly the initial longitudinal positions of jet production along the jet direction \cite{Du:2020pmp,Du:2021pqa}.

The above methods in previous studies can only provide very primitive estimate of the averaged initial jet production positions. To fully understand the jet-induced  medium response and look for signals of Mach cones and the diffusion wake, more accurate and event-by-event jet production positions are desired, taking into account more realistic jet energy loss and the fluctuating nature of jet shower development in the QGP medium.
We aim to develop a deep learning assisted jet tomography in this study that will address the following questions:
\begin{enumerate}[itemsep=-.4em]
    \item Is it possible to determine the initial jet production positions using the momenta of final state particles?
    \item Will the signal of Mach cones and the diffusion wake be amplified if initial jet production positions are constrained to specific regions with long jet propagation lengths or fixed propagation direction relative to the radial flow? 
    \item How reliable is the new deep learning assisted method?
\end{enumerate}
To answer the first question, we employ a point-cloud deep neural network  (DNN) with multiple hidden layers, which is powerful in pattern recognition and has been widely used in high energy nuclear physics \cite{Pang:2021vwl,Larkoski:2017jix,RevModPhys.91.045002,Boehnlein:2021eym} and jet medium interaction studies \cite{Apolinario:2021olp,Lai:2021ckt,Ying:2022jvy,Brewer:2021hmh,Chen:2019apv,Chien:2018rgm,Brewer:2020och,Du:2020pmp,Du:2021pqa,Liu:2022hzd}. We will train and validate the DNN using data from  the coupled Linear Boltzmann Transport (CoLBT) and hydro model simulations \cite{Chen:2017zte} which combines the Linear Boltzmann Transport (LBT) model for jet parton propagation in medium with the concurrent QGP evolution given by the 3+1D CCNU-LBNL viscous (CLVisc) hydrodynamics \cite{Pang:2012he,Pang:2014ipa,Pang:2018zzo}. The CoLBT-hydro model also simulates the stochastic energy momentum deposition to the medium by jet-shower partons together with the relativistic hydrodynamic medium response. For the second question, we can average events with the same jet production regions as predicted by deep neural network.
This method, dubbed ``jet-position engineering'', shows that the signal of Mach cones and diffusion wake depends strongly on the initial jet production regions.
The third question concerns with the training model dependence of the neural network. We will use pseudo-data generated from another model, the LIDO partonic transport model \cite{Ke:2018jem, Ke:2020clc}, to cross-check performance of the trained network, i.e., use the CoLBT-trained network to infer the jet production locations from the pseudo-data generated from the LIDO model and compare to the ground truth. Although we do not expect either model to provide the exact real jet dynamics in the QGP, this practice can estimate the model uncertainty of the method.

The remainder of the paper is organized as following. In Section II, we describe the major physics model CoLBT-hydro and the cross-check model LIDO for jet evolution in the QGP. We will also introduce the point-cloud neural network method for locating the initial jet production positions from the information of final state hadrons.
Results are presented in Section III. Section IV summarizes the paper.

\section{Method}

{\it CoLBT-hydro model for jet transport and medium response --} The initial configurations in momentum distributions of $\gamma$-jets are generated by PYTHIA8~\cite{Sjostrand:2007gs}. The Trento model is used for both the initial spatial distribution \cite{Moreland:2014oya} of $\gamma$-jet production and the initial energy density distribution for the hydrodynamic evolution of the bulk QGP medium. The jet transport in QGP is simulated using the LBT model~\cite{Wang:2013cia,He:2015pra,Cao:2016gvr,Xing:2019xae},
where evolution of the phase space distribution of jet shower partons in QGP is described by the linearized Boltzmann equation,
\begin{equation}
p_a\cdot\partial f_a = \int \sum_{b c d } \prod_{i=b,c,d}\frac{d^3p_i}{2E_i(2\pi)^3} (f_c f_d-f_a f_b)|{\cal M}_{ab\rightarrow cd}|^2 \frac{\gamma_b}{2}
S_2(\hat s,\hat t,\hat u)(2\pi)^4\delta^4(p_a\!+\!p_b\!-\!p_c\!-\!p_d)+ {\rm inelastic},
\label{bteq}
\end{equation}
where $p_{a/c}$ and $f_{a/c}$ are the four-momenta and phase space distributions of jet-shower partons before and after each scattering, $p_{b/d}$ are the four-momenta of medium partons sampled from thermal distributions $f_{i} = 1/(e^{p_{i}\cdot u / T} \pm 1)$ ($i=b, d$) in a locally thermalized QGP with temperature $T$ and fluid four-velocity $u$. The left hand side of the equation describes the change of parton phase space density $f_a$. The right hand side contains the loss term ($-f_a f_b$) and the gain term ($f_c f_d$). The loss term describes the forward process where a jet shower parton $a$ in the given phase space scatters with a thermal parton $b$, causing a decrease in $f_a$ if parton $a$ leaves this specific phase space. The gain term describes the back reaction, where a jet shower parton $c$ scatters with a thermal parton $d$, causing an increase of phase space density $f_a$, if the recoiled parton enters the specific phase space. Both processes have considered all possible channels of elastic scattering $a + b \rightarrow c + d$, whose amplitudes $|{\cal M}_{ab \rightarrow cd}|$ are computed up to leading order in perturbative QCD \cite{Eichten:1984eu}. The factor $\gamma_b$ is the color and spin degeneracy of parton $b$ in the forward process and
\begin{equation}
S_2(\hat s, \hat t, \hat u) = \theta(\hat s\ge 2\mu_{D}^2)\theta(-\hat s+\mu_{D}^2\le \hat t\le -\mu_{D}^2)
\label{eq:cutLBT}
\end{equation}
is a Lorentz-invariant double step function used to regularize the collinear divergence in the scattering amplitudes, where $\hat s$, $\hat t$, and $\hat u$ are the Mandelstam variables, and $\mu_{D}^2 = \frac{3}{2}g^2 T^2$ is the Debye screening mass. The inelastic part in Eq.~\eqref{bteq} is the contribution from gluon radiation that is induced by the elastic scattering processes. The induced gluon emission rate $\Gamma^{\rm inel}_{a}$ is calculated according to the higher-twist formalism \cite{Guo:2000nz,Majumder:2009ge,Zhang:2003wk}. Please refer to Refs.~\cite{Wang:2013cia,He:2015pra,Cao:2016gvr,Xing:2019xae} and  ~\cite{Luo:2018pto,He:2018xjv,Chen:2017zte,Chen:2020tbl} for more details about LBT and CoLBT-hydro model.

The four-momenta of thermal partons $b$ and $d$ are sampled from distribution functions 
$f(p) =  1/(e^{p\cdot u / T} \pm 1)$ with $+$ for quarks and $-$ for gluons. 
The local temperature and fluid velocity $u$ are provided by the CLVisc model that solves numerically relativistic hydrodynamic equations, 
\begin{equation}
    \nabla_{\mu} T^{\mu\nu} = J^{\nu},
\end{equation}
for the space-time evolution of QGP, where $\nabla_{\mu}$ is the covariant derivative operator, $T^{\mu\nu} = (\epsilon+P)u^{\mu}u^{\nu} - P g^{\mu\nu} + \pi^{\mu\nu}$ is the energy stress tensor, in which $\epsilon$ and $P$ are the energy density and pressure in the co-moving frame of the fluid, $u^{\mu}$ is the relativistic fluid four-velocity satisfying $u_{\mu}u^{\mu} = 1$, $g^{\mu\nu} = \rm{diag}(1, -1, -1, -\tau^2)$ is the metric tensor in the Milne $(\tau, x, y, \eta_s)$  coordinates and $\pi^{\mu\nu}$ is the shear-stress tensor. The source term $J^{\nu}$ describes the energy and momentum deposition from jet showers to QGP. It includes the energy and momentum from the recoiled and radiated partons  whose  energy in the local co-moving frame of the fluid is smaller than a cut-off which is set to $p_0=2$ GeV/$c$.  Partons taken out of the phase space in the medium will leave a particle hole behind, and will deposit negative energy-momentum to the medium through the source term. This source term $J^{\nu}$ is responsible for generating the Mach cones and the diffusion wake in the QGP medium. This group of time-dependent partial differential equations are solved numerically on GPU using OpenCL in the CLVisc model concurrently with the LBT transport. See Refs.~\cite{Pang:2012he,Pang:2014ipa,Pang:2018zzo,Wu:2021fjf} for more details. A parton recombination model is used to hadronize both the QGP and hard partons from jet showers~\cite{Zhao:2020wcd,Zhao:2021vmu,Han:2016uhh}. The kinetic cuts in this work are $p_T^\gamma\in(200,250)$ GeV, $\left | \eta_\gamma \right |<0.32$, $p_T^{\rm jet}>100$ GeV, $\left | \eta_{\rm jet} \right |<0.75$, respectively, with jet cone size $R=0.4$.

{\it LIDO transport model for jet evolution.}~The LIDO model also contains the essential ingredients of collisional and radiative processes as well as the medium response, but differs from CoLBT-hydro in details.
We will apply the predictor trained with the CoLBT-hydro model to infer the initial jet production locations from a set of pseudo-data generated by the LIDO transport model. This provides an estimation of the model-dependent uncertainty in the predictor. Here, we briefly describe the LIDO model with an emphasis on its difference as compared to the CoLBT-hydro model. Full details of the LIDO model can be found in Refs.~\cite{Ke:2018jem, Ke:2020clc}.

The evolution of the distribution function of hard partons in the LIDO model can be schematically written as
\begin{eqnarray}
\frac{d f(t, x, p)}{dt} = \left[-\eta \nabla_p+ \frac{1}{4}\hat{q}_s\nabla_p^2\right]f+\mathcal{C}_{12}[f] + \mathcal{C}_{22}[f] + \mathcal{C}_{23}[f] .
\end{eqnarray}
The first key difference as compared to the CoLBT-hydro model is that the jet-medium interaction is separated into two parts: 1) hard interactions with $q_\perp > 2\mu_D$ are simulated according to the matrix elements of scatterings in the vacuum $\mathcal{C}_{22}[f]$; 2) soft interactions that transfer small momentum $q_\perp <2\mu_D$ between jet partons and the medium are modeled by a diffusion process, as indicated by the first term on the right-hand side of the equation. The soft diffusion contant is $\hat{q}_s = \alpha_s C_R \mu_D^2 T \ln 4$ with the corresponding soft drag constant $\eta = \hat{q}_s/4ET$. Since a large fraction of the elastic processes are taken care of by a diffusion equation (Langevin dynamics in particle-based simualtions), the fluctuation of elastic energy loss in LIDO is smaller than in CoLBT-hydro with $-t\approx q_\perp^2 > \mu_D^2$. 
The second difference is the handling of radiative processes. The LIDO model splits parton emissions into diffusion-induced radiation $\mathcal{C}_{12}[f]$ and hard-collision-induced radiations $\mathcal{C}_{23}[f]$. The LIDO model applies a different implementation of the Landau-Pomeranchuck-Migdal (LPM) interference effects that works best in the deep LPM region, while the CoLBT-hydro model applies the Higher-twist formula \cite{Guo:2000nz}. Nevertheless, both methods result in a radiative energy loss that scales as $\ln E$ for energteic partons.

The parameters in CoLBT-hydro and LIDO model have been calibrated to reproduce experimental data on the suppression of single inclusive jet and hadron spectra \cite{He:2018xjv,Zhao:2021vmu,Ke:2020clc}. Therefore, both models have similar averaged jet/parton energy loss, while the fluctuation and traverse dynamics could be different. This will be elaborated later in the next section where we compare hadron distributions within jets between the two models.

Finally, the LIDO model does not solve the fully coupled hydrodynamic response, but uses a simplified formula to compute the correction to jet energy from medium response \cite{Ke:2020clc}. Furthermore, LIDO only uses Lund string fragmentation  while the CoLBT-hydro model uses fragmentation plus parton recombination model for hadronization. Nevertheless, we expect these differences only affect particle distribution at low $p_T$. We do not consider these to have major impacts on the following inference of jet locations, where only hadrons with $p_T>2$ GeV/$c$ are used in the training.

\begin{figure}[htbp]
    \centering
    \includegraphics[width=0.85\textwidth]{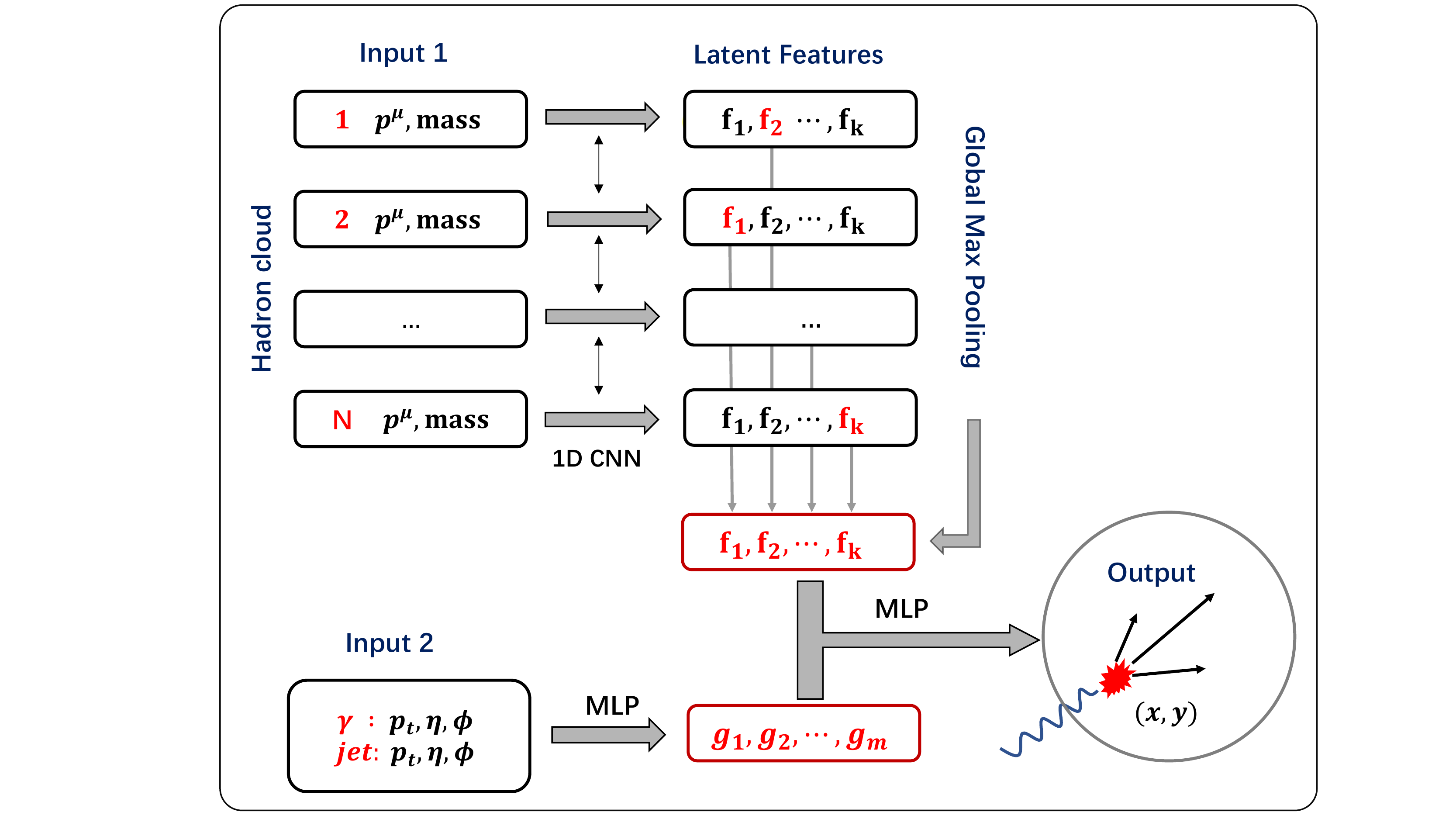}
    \caption{(color online) The structure of the point cloud network that is used to predict the jet production positions $(x, y)$ from 
    hadrons in the jet and the global information of $\gamma$ and jet.}
    \label{fig:network}
\end{figure}

{\it Point Cloud Network for particles in momentum space} We employ a point cloud network to locate the initial jet production positions as shown in Fig.~\ref{fig:network}.
The network has two kinds of inputs for each training sample.
The first input consists of a list of hadrons in the jet with 6 features ($p_x$, $p_y$, $p_z$, $E$, $mass$, $p_T$) for each hadron.
Since different events may have different numbers of hadrons,
we concatenate a list of zeros to the end of the hadron list to use the point cloud network. Hadrons from jets are selected with $p_T>2$ GeV/$c$ to reduce the influence of soft hadrons from the hydrodynamic background.
The second input contains the global information about the $\gamma$-jet event,
with 6 floating numbers indicating the ($\eta, p_T, \phi$) of the $\gamma$ and
the jet reconstructed using the FastJet with anti-kt algorithm \cite{Cacciari:2011ma}.

The point cloud network uses one dimensional convolution neural network (1DCNN) to 
convert the 6 features of each hadron in the cloud to high dimensional latent space.
The 1DCNN is a feed forward neural network (also called multi-layer perception or MLP) whose parameters are shared by all hadrons in the cloud.
The 1DCNN has a structure of  MLP(256, 128, 128) where the numbers stand for the number of neurons in each hidden layer. At each step, an ELU activation function is used to get non-linear transformation capability. The high dimensional latent features of each hadron from 1DCNN are independent of each other. Global Max Pooling will locate the boundaries along each direction in the high dimensional latent space. This step preserves the permutation symmetry and helps to capture the correlations between hadrons in the jet. The output of this step is the jet sub-structure information encoded in 128 numbers from the Global Max Pooling. Another MLP(256, 128) is used to convert the second input to high dimensional features including $\gamma$-jet asymmetry. Latent features from these two inputs, corresponding to local and global information, are fed together to the final MLP(32) to produce the output. The output of the network has two neurons representing the initial jet production position $(x^{\rm net}_i, y^{\rm net}_i)$ in the transverse plane, for given sets of the final hadron four-momenta $\{\vec{p}\}_i$,
\begin{align}
(x^{\rm net}_i, y^{\rm net}_i) = f(\{\vec{p}\}_i, \theta),
\end{align}
where $f$ is the function represented by the neural network, $\theta$ is the collection of all the trainable parameters which are initialized with random numbers.
The training objective is to minimize the loss which is the squared differences between network prediction and the ground truth,
\begin{align}
    {\rm loss} = {1\over N}\sum_{i=1}^N \left[ (x_i^{\rm net} - x_i^{\rm true})^2 + (y_i^{\rm net} - y_i^{\rm true})^2 \right],
\end{align}
where $N$ is the total number of hadrons from jets in a mini-batch of training samples.
The loss is reduced step-by-step using the stochastic gradient descent (SGD),
\begin{align}
 \theta^{\rm new} = \theta^{\rm old} - \alpha {\partial {\rm loss} \over \partial \theta},
\end{align}
where the learning rate $\alpha$ is a small positive number. 
In this study, we use the Adam optimizer which improves SGD using momentum mechanism and adaptive learning rate.

\begin{figure}
    \centering
    \includegraphics[width=0.6\textwidth]{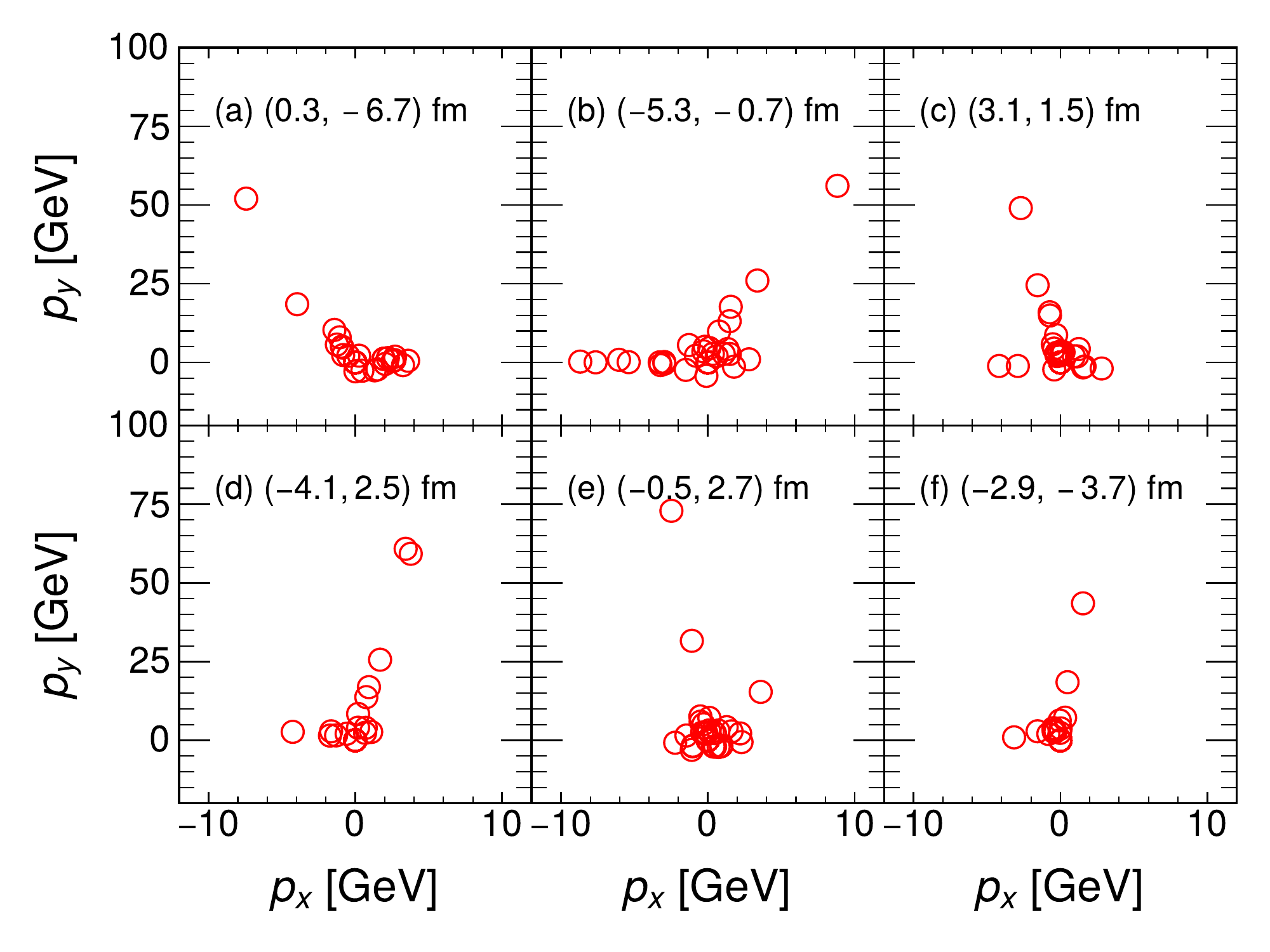}
    \caption{(color online) The $p_x$ and $p_y$ (circles) of jet hadrons for 6 randomly selected $\gamma$-jet events and their corresponding jet production positions $(x, y)$ in the transverse plane.}
    \label{fig:inputs}
\end{figure}

Shown in Fig.~\ref{fig:inputs} are $p_x$ and $p_y$ (circles) of jet hadrons for 6 randomly selected CoLBT-hydro events with a $\gamma$-jet and their corresponding jet production positions $(x, y)$ in the transverse plane. These six scatter plots illustrate the training data and the corresponding labels used in the supervised regression. No obvious patterns in the $(p_x, p_y)$ distributions of jet hadrons can be found visually that are directly related to the jet production positions due to the stochastic nature of jet showering and scattering in the medium. 
This reminds us that the neural network is not expected to determine the jet production positions accurately in all events, if the causality between the input and output is destroyed by event-by-event fluctuations.

The network prefers to work with numbers between -1 and 1.
During training, the four-momenta and mass of hadrons are scaled-down by a factor of 2000.
The transverse momentum $p_T$ of $\gamma$ and jet are scaled-down by a factor of 250.
The coordinates x and y of initial jet production are scaled-down by a factor of 10.
Because of the scaling, a typical loss $0.06$ correspond to an average deviation $\sqrt{6}=2.4$ fm from the ground truth. 

\begin{figure}[!htp]
    \centering
    \includegraphics[width=0.5\textwidth]{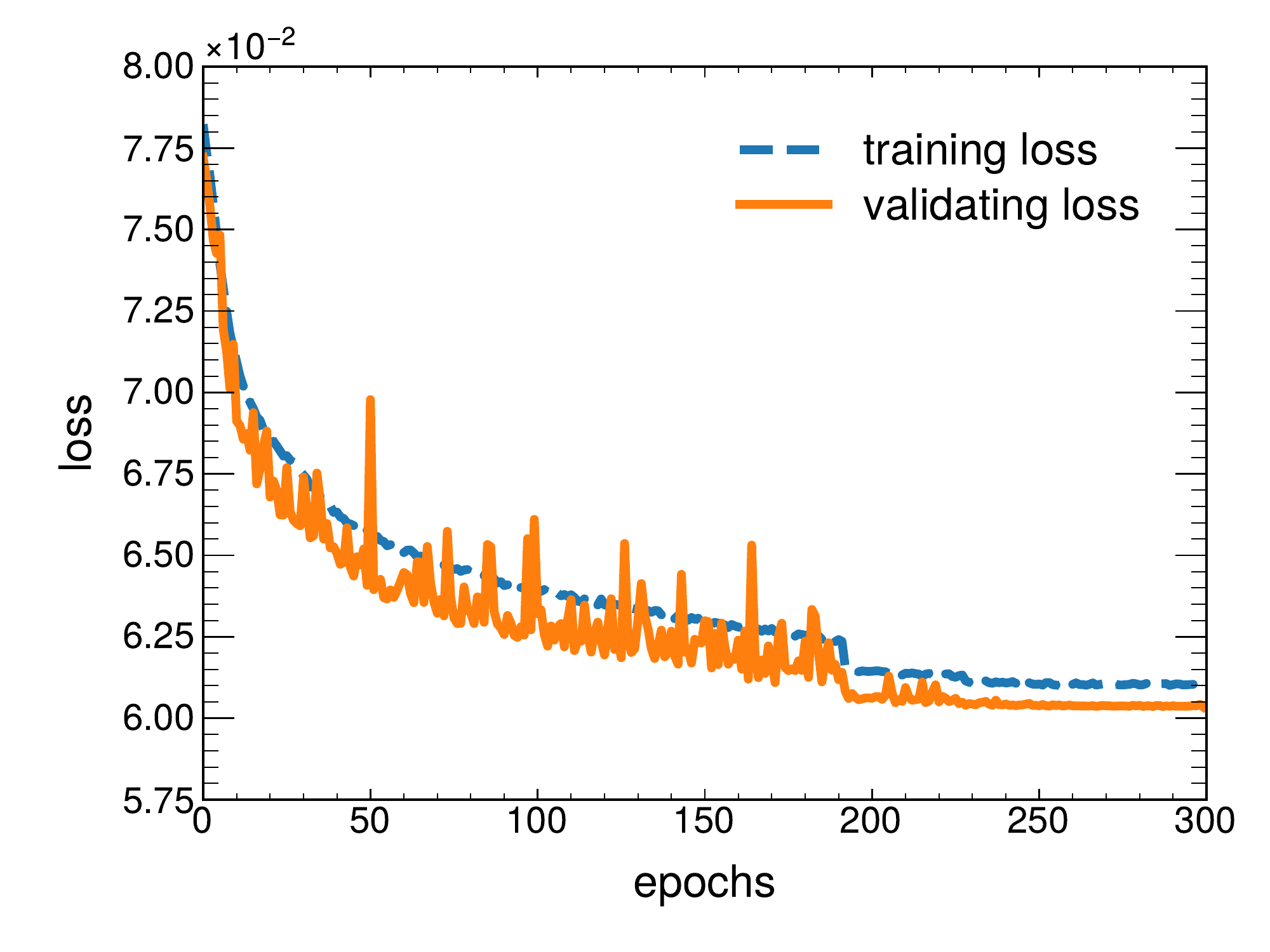}\includegraphics[width=0.5\textwidth]{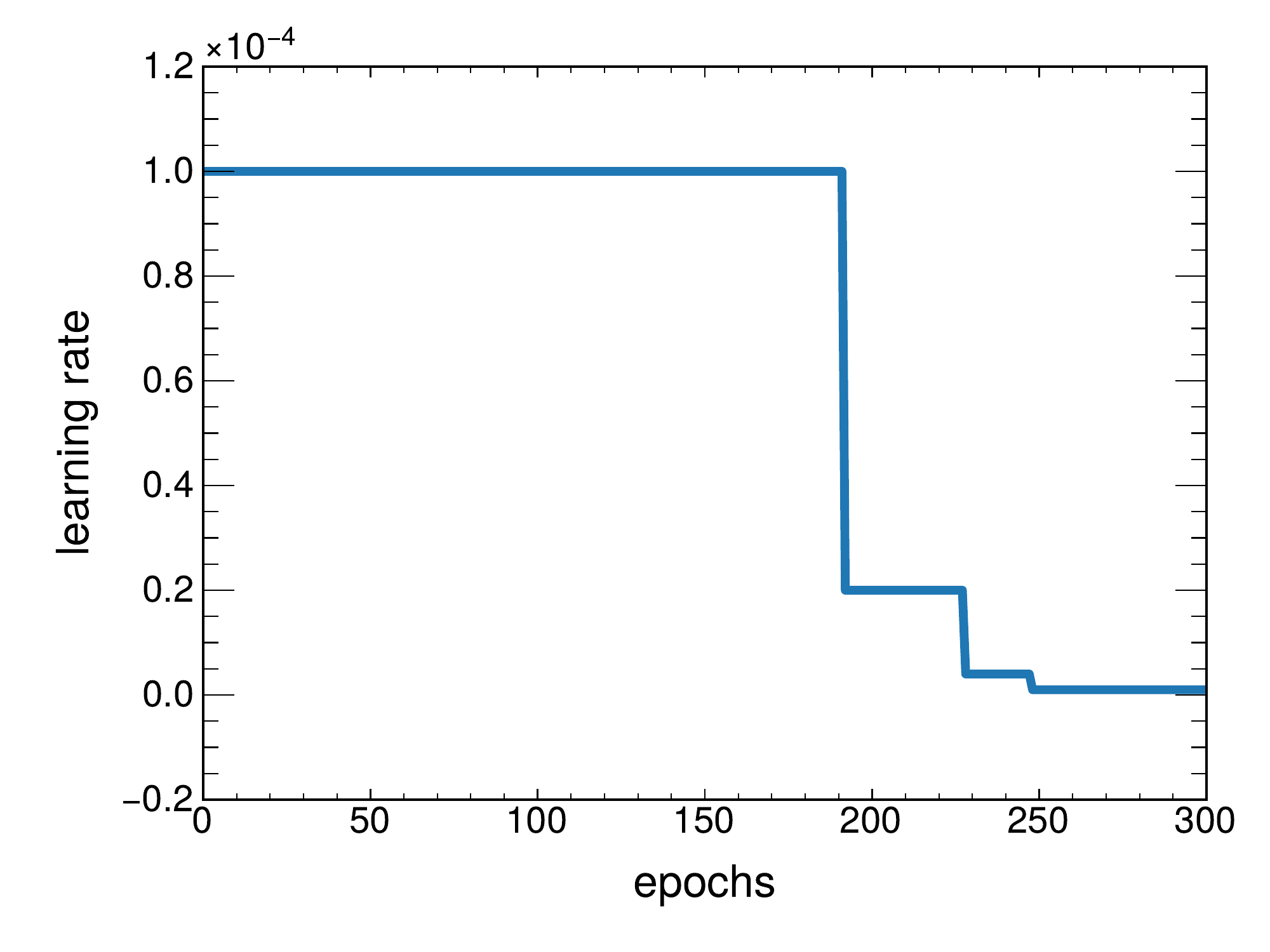}
    \caption{(color online) The training history as a function of epochs. Left figure shows the training and validating loss of the network. Right figure shows the adaptive learning rate changing with the training process. The learning rate is a key parameter in stochastic gradient descent. A large learning rate at the beginning of the training helps to walk fast in the parameter space, while a small learning rate at late time will reduce oscillations around the minima and helps the network to converge.}
    \label{fig:train_history}
\end{figure}

{\it Locating jet production positions using deep learning} 
Fig.~\ref{fig:train_history} shows the training and validating loss as a function of training epochs in the jet-production-position predictions.
The first epoch is trained with batch-size 4 which is different from other epochs whose batch-size equals 32. By doing this, the network can explore more regions first and then continues to optimize from a good starting point. Besides the adaptive learning rate provided by the Adam optimizer, we reduce the learning rate by 20\% if the validation loss saturates and stays on a plateau for 50 epochs.
The validating and testing errors using data from CoLBT-hydro are about 0.06 which correspond to a root-mean-square-error (RMSE) $\sqrt{6}=2.4$ fm.

Using pseudo-data from LIDO model simulations to test our trained neural network,
we observe a smaller testing error than the validating and testing error on CoLBT-hydro data. The testing error is only 0.049 on data from LIDO model, which corresponds to a RMSE 2.2 fm. In the first glance, it seems mysterious since usually the testing error is larger than training error unless there are dropouts which randomly discard a fraction of neurons during training to increase the generalization ability.
In our case, we did not use dropout and suspect that the testing error is smaller on LIDO data because of less fluctuations in the LIDO simulations.

\begin{figure}[htbp]
    \centering
    \includegraphics[width=0.6\textwidth]{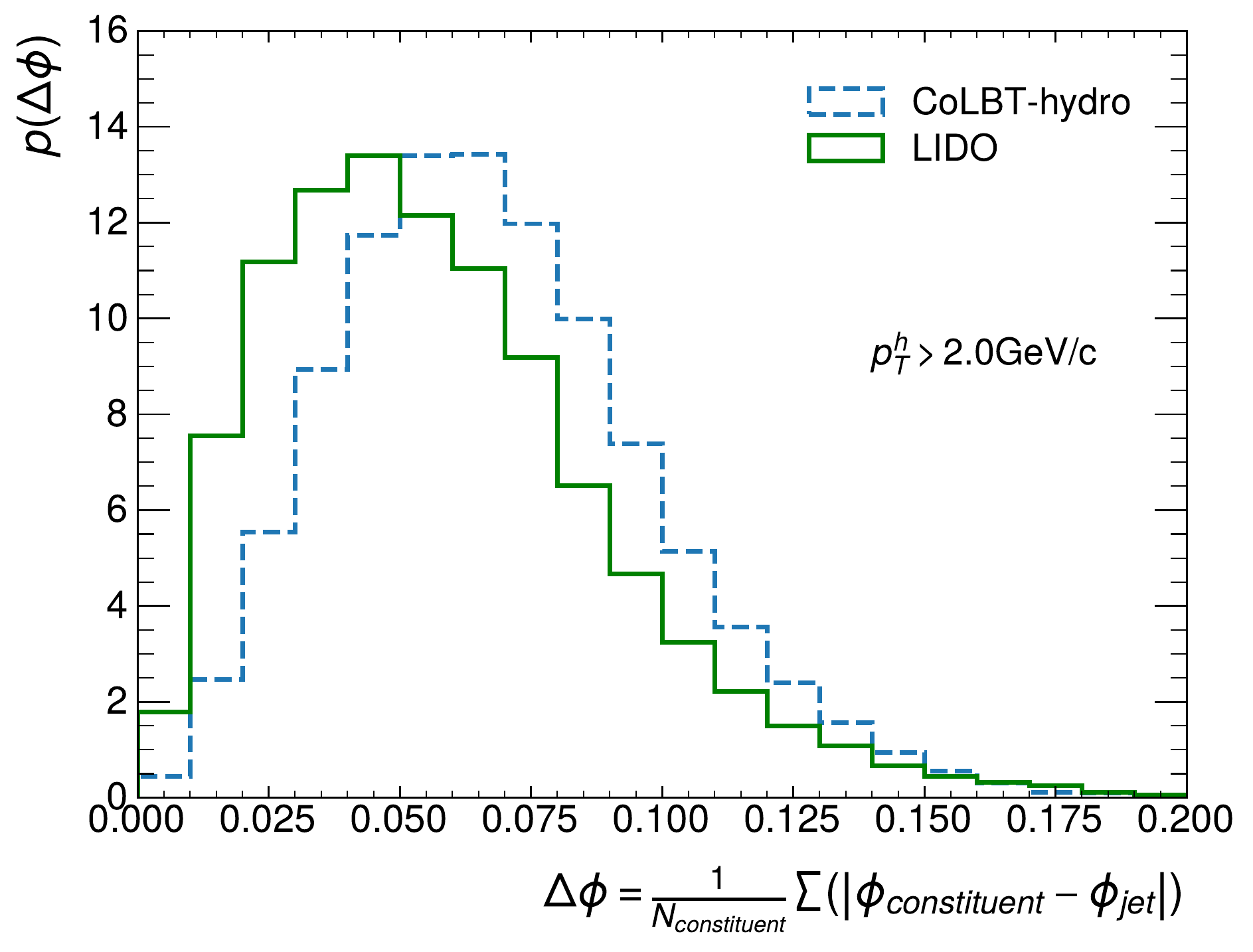}
    \caption{(color online) The distribution of mean $\Delta\phi$ for hadrons in jet. The green solid line is from LIDO  model and the blue dashed line is from CoLBT-hydro model }
    \label{fig:lido_vs_lbt}
\end{figure}

One notable difference is that CoLBT-hydro uses both fragmentation and recombination for hadronization while LIDO uses fragmentation only. To check the effect of this model difference, we switch off the recombination in CoLBT-hydro and found that the testing error is reduced from 0.060 to 0.054.

There are other model differences which affect the jet shape and the predicting accuracy. Fig.~\ref{fig:lido_vs_lbt} shows the distribution of the mean absolute angle of hadrons in jet, with respect to the jet-axis.
Hadrons spread to larger angles in CoLBT-hydro as compared with LIDO.
Both models have taken into account the rescattering of recoiled partons.
LIDO considers the $t$-channel scattering between partons,
while CoLBT-hydro has also $s$ and $u$-channels, which result in more wide-angle scatterings. Other model differences between CoLBT-hydro and LIDO might also contribute to this surprisingly better prediction accuracy on pseudo-data from LIDO whose underlying modeling is different from the CoLBT-hydro model used for training data production.

\section{Results}

\begin{figure}[htbp]
    \centering
    \includegraphics[width=1.0\textwidth]{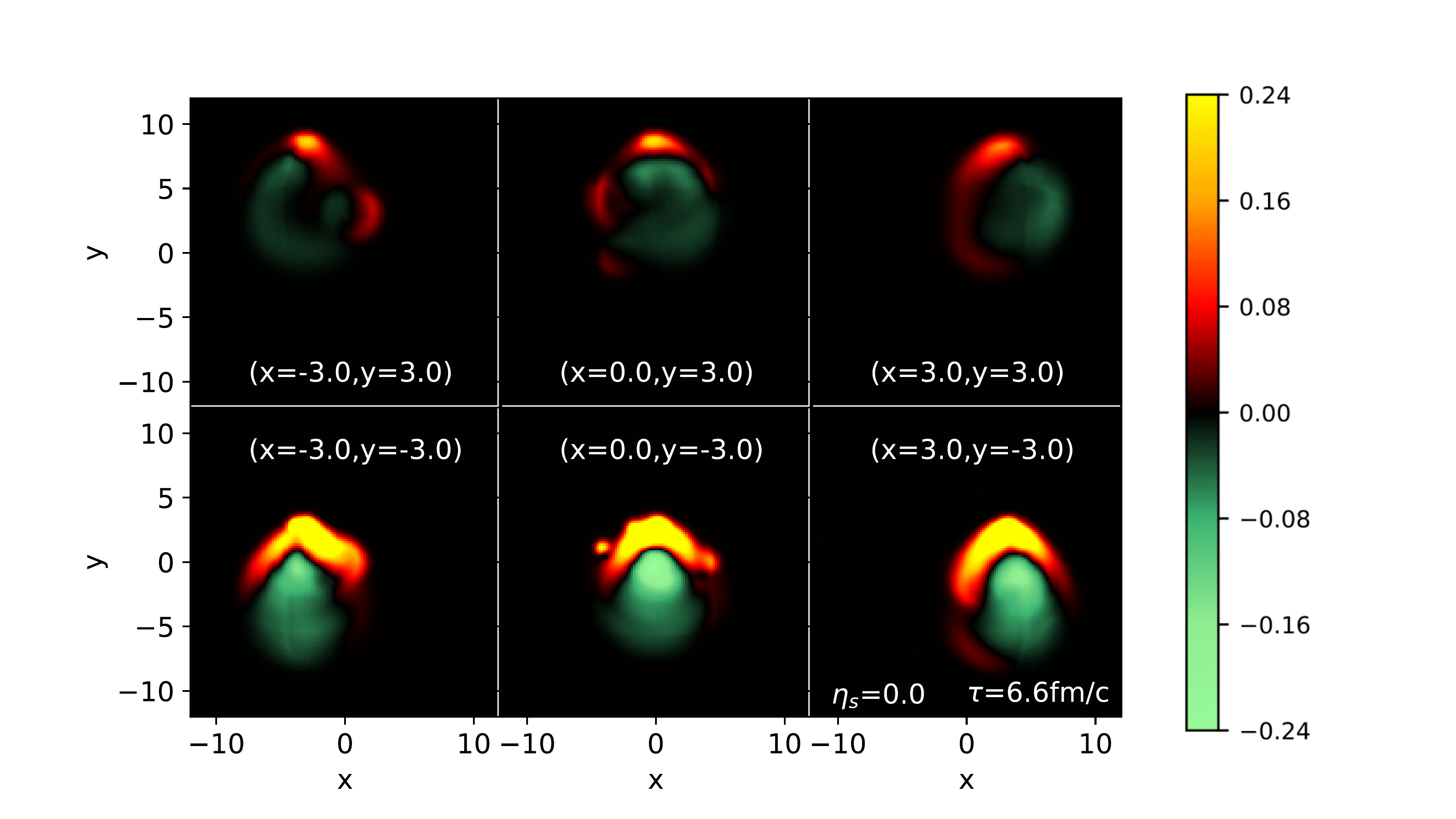}
    \caption{(color online) Energy density transverse distributions of jet-induced medium response from CoLBT-hydro simulations of 6 $\gamma$-jet events with $p_T^\gamma \in(200,250)$ GeV/$c$ in  0-10\% central Pb+Pb collisions at $\sqrt{s}=5.02$ TeV with different jet production positions.}
    \label{fig:demo}
\end{figure}

{\it The Mach cones from CoLBT--} 
Before we study signals of jet-induced Mach cones and diffusion wake with specified spatial regions of the initial jet production positions using the trained point cloud neural network, we first look at features of the Mach cones from CoLBT-hydro simulations and their dependence on the initial jet production positions. Shown in Fig.~\ref{fig:demo} are the transverse distributions of the energy density at spatial rapidity $\eta_s=$0.0 from jet-induced medium response in a single $\gamma$-jet event, with 6 different initial jet production positions. Jets propagate upward in all these events. In this plot background from the same hydrodynamics events without jets is subtracted. The negative energy density presents the depletion of the medium due to the diffusion wake. The upper panels show events where jets are initially produced close to the upper edge of the overlap region of two colliding nuclei and travel outward. These jets in general have short path lengths and lose less energy. The lower panels show events where jets are produced near the lower edge of the overlap region but travel inward through the center and across the whole length of the QGP with long path lengths. The average medium density (temperature) on their paths is also higher.  The left and right panels show events in which the jet propagation direction is tangential to the radial flow while the middle panels show events where jets travel along (upper-center) and against (lower-center) the radial flow.

The medium responses and signals of the Mach cones are clearly stronger for longer path lengths and higher average medium density on the jet propagation path. The shape of the Mach cone is also apparently distorted by the temperature gradient and the radial flow of the QGP medium. For jets traveling tangentially to the radial flow (left and right panels), jet shower partons going toward the center of QGP lose more energy and induce stronger medium response as compared to these partons going away from the center. On the other hand, momenta of the thermal partons associated with the flowing medium will blow the medium response along the direction of the radial flow as seen in the shape of the diffusion wake which is tilted in the direction of the radial flow. The final distorted shape of the Mach cone is the result of competition between temperature gradient and radial flow. This distorted pattern is clearly seen in the left and right panels of Fig.~\ref{fig:demo} where both the Mach cone and the diffusion wake are asymmetrical with respect to the jet propagation (or $\gamma$) direction. Observation of this asymmetric or distorted medium-response in the final hadron distributions with different jet production positions will provide clear evidence of jet induced Mach cones. In the following we will demonstrate that one can use a trained deep neural network to locate the jet initial production positions from the full information of final state hadrons inside the jet-cone. Soft-hadron distributions in events classified by the neural network according to the initial jet production positions should show the expected asymmetry.

\begin{figure}[!htp]
    \centering
    \includegraphics[width=0.95\textwidth]{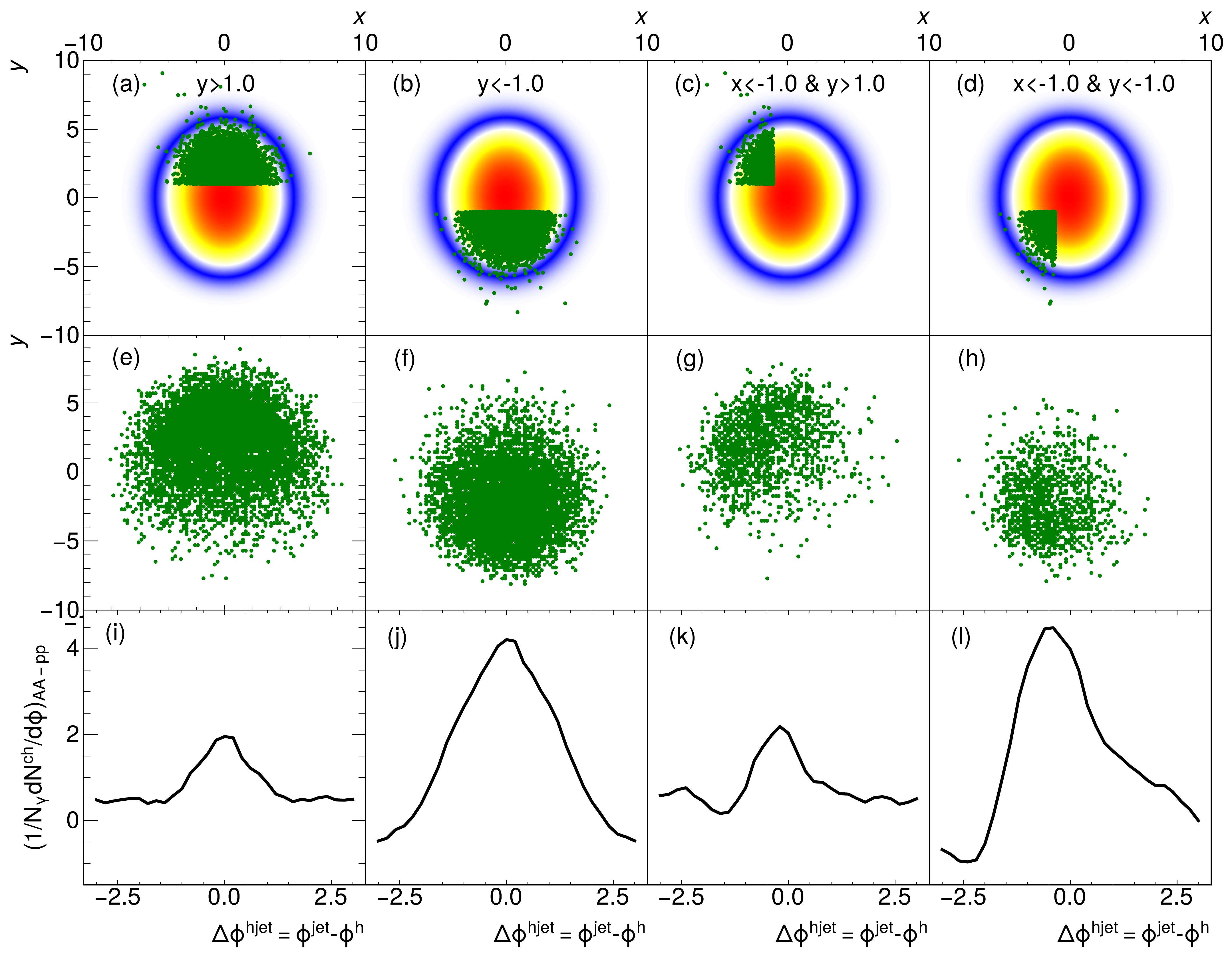}
    \caption{(color online) The initial jet positions in 0-10\% central Pb+Pb collisions selected by the deep neural network (upper panels) as compared to the true initial jet production points in CoLBT-hydro simulations (middle panels) and the corresponding difference in $\gamma$-hadron yields as a function of $\Delta \phi^{\rm h jet }$=$\phi^{\rm jet}-\phi^{\rm h}$ between Pb+Pb and p+p collisions (lower panels). See text for more details.} 
    \label{fig:dndphi}
\end{figure}

{\it Jet-hadron correlation with engineered initial jet production positions--}
We can now use the trained neural network to classify $\gamma$-jet events from CoLBT-hydro simulations according to the predicted positions of the initial jet production from the neural network and look at the corresponding $\gamma$-hadron (or jet-hadron) correlations. Shown in Fig.~\ref{fig:dndphi} are scatter plots of jet production positions selected by the deep neural network (upper panel), the true production positions in the simulated events (middle panel) and the corresponding jet-hadron correlations in azimuthal angle for charged hadrons (lower panel) in 0-10$\%$ Pb+Pb collisions at the colliding energy $\sqrt{s}=5.2$ TeV. The baseline correlation in proton+proton collisions is subtracted. The transverse momenta of the prompt photon and jets are $p_T^\gamma\in(200,250)$ GeV and $p_T^{\rm jet}>100$ GeV, respectively with jet cone size $R=0.4$.  Their rapidities lie in $\left |\eta_\gamma \right |<0.32$  and $\left |\eta_{\rm jet}\right |<0.75$. The transverse momenta of soft charged hadrons are restricted to $p_T^h\in(1.0,2.0)$ GeV.
The contour plots in the upper panels show the averaged temperature profile of the QGP background aligned to the reaction plane of the collisions. The photon($\gamma$) is along the negative y direction and jet shower partons traverse in the opposite direction.

It is not possible to determine the jet production position exactly in each event due to large fluctuations even with the complete information of jet hadrons. For the same reason it is also difficult to locate the average jet production points precisely using the longitudinal and transverse gradient tomography. The neural network predictions are also approximate as one can see here. Even though the true initial production points spread to a larger region than the network predictions, clipping the network predicted jet production points into different sections will still localize the true production positions to the region of interests as one can see in comparing the upper and middle panels of Fig.~\ref{fig:dndphi}.

For the network selected regions of initial jet production points in the right [Figs.~\ref{fig:dndphi}-(d, h, l)] and center-left panels [Figs.~\ref{fig:dndphi}-(b, f, j)], the $\gamma$-jet has to traverse (in the positive $y$-direction) the central region of the QGP medium with a longer path length where the temperature is also higher. This causes more jet energy loss and leads to a larger yield of soft hadrons from the medium response and induced radiation as compared to the cases in the left [Figs.~\ref{fig:dndphi}-(a, e, i)] and center-right panels [Figs.~\ref{fig:dndphi}-(c, g, k)] where jets are produced in the outer region and travel away from the center of the QGP with a shorter path length where the temperature is also lower. We should note that the enhanced soft hadrons in the peak region along the jet direction shown in the lower panels of Fig.~\ref{fig:dndphi} can come from both the Mach cone and medium-induced gluon radiation. It is difficult to distinguish these two sources of soft hadron enhancement in the measured distributions. 

As one can see in Fig.~\ref{fig:demo}, the diffusion wake associated with jet-induced Mach cones is just a depletion of the hydrodynamic background behind the propagating jet. This will lead to a depletion of soft hadrons in the $\gamma/Z^0$ direction in $\gamma/Z^0$-jet events. One can clearly see this depletion (yields are negative after subtraction of the p+p baseline) in Figs.~\ref{fig:dndphi}(j) and (l) where the effect of the medium response is strong. As pointed out in Ref.~\cite{Yang:2021qtl}, mini-jets from multiple parton interactions (MPI) associated with the hard processes of $\gamma/Z^0$-jet production contribute to a background in the jet-hadron correlation that is constant in azimuthal angle. These mini-jets will also interact with the QGP medium and experience jet quenching in heavy-ion collisions. Quenching of these MPI mini-jets will lead to an enhancement of soft hadrons from the MPI background some of which still remain even after the subtraction of p+p baseline. This is the reason why one does not see the clear depletion due to diffusion wake in Fig.~\ref{fig:dndphi}(i) where the effect of the medium response is weak. One can use a mixed-event method to estimate the enhanced MPI background in A+A collisions. The depletion of soft hadrons due to the diffusion wake will become stronger after the subtraction of the MPI background \cite{Yang:2021qtl}.

As we have illustrated in Fig.~\ref{fig:demo}, the shape of Mach cones and diffusion wake are distorted both by the radial flow and the density gradient of the expanding QGP. They also influence the jet-hadron correlation in the final soft hadrons. When the distribution of the initial jet production positions is symmetric with respect to the jet propagation (or $\gamma$) direction, the jet-hadron correlation should also be symmetric with respect to the jet direction ($\Delta\phi=0$) as shown in Figs.~\ref{fig:dndphi}(a,e,i) and (b,f,j). In these symmetric jet-hadron correlations, the width of the correlation is broadened by both the wind of the radial flow and the slope of the density gradient. The broadening is larger for longer jet path lengths. The depletion due to the diffusion wake peaks in the direction of the trigger $\gamma$ ($\Delta\phi=\pi$) as seen in  Fig.~\ref{fig:dndphi}(j).

If the distribution of the initial jet production position is not symmetric with respect to the jet direction,  jet propagation is tangential to the radial flow most of the time. The Mach cones and diffusion wake will be distorted by the temperature gradient and radial flow and 
become asymmetric with respect to the jet direction as seen in Fig.~\ref{fig:demo}.
This asymmetry should also be reflected in the soft hadron azimuthal angle distributions as seen in the right two panels of Fig.~\ref{fig:dndphi} [(e,g,k) and (d,h,l)] for $\gamma$-jet events in which the initial jet production locations are selected by the neural network such that jet propagation is mostly tangential to the radial flow.  In these asymmetric jet-hadron angular correlations,  hadron yields in the direction towards the central region of the QGP are enhanced because of the higher temperature of the medium, larger parton energy loss and stronger medium response.  The diffusion wake is, on the other hand, both swiped by the radial flow and deflected by the density gradient such that the minimum angle (dip in the soft hadron depletion) is shifted away from the $\gamma$ direction $\Delta\phi^{\rm hjet}=\pi$. The angular shift is larger when the diffusion wake is exposed more to the radial flow and density gradient as in the case of Figs.~\ref{fig:dndphi}(e,g,k). The dip is shallower due to smaller jet path length and weaker medium response as compared to the case in Figs.~\ref{fig:dndphi}(d,h,l). Similarly, the position of the peak in the jet-hadron angular correlation is also shifted away from the jet direction ($\Delta\phi=0$) due to the influence of the radial flow and the density gradient. The shift is larger when the jet propagation path length is longer. The asymmetrical features in the measured jet-hadron correlations in events with neural network engineered jet production locations should be the clear evidence of the jet-induced Mach cones and diffusion wake in heavy-ion collisions.

\begin{figure}[!htp]
    \centering
    \includegraphics[width=0.45\textwidth]{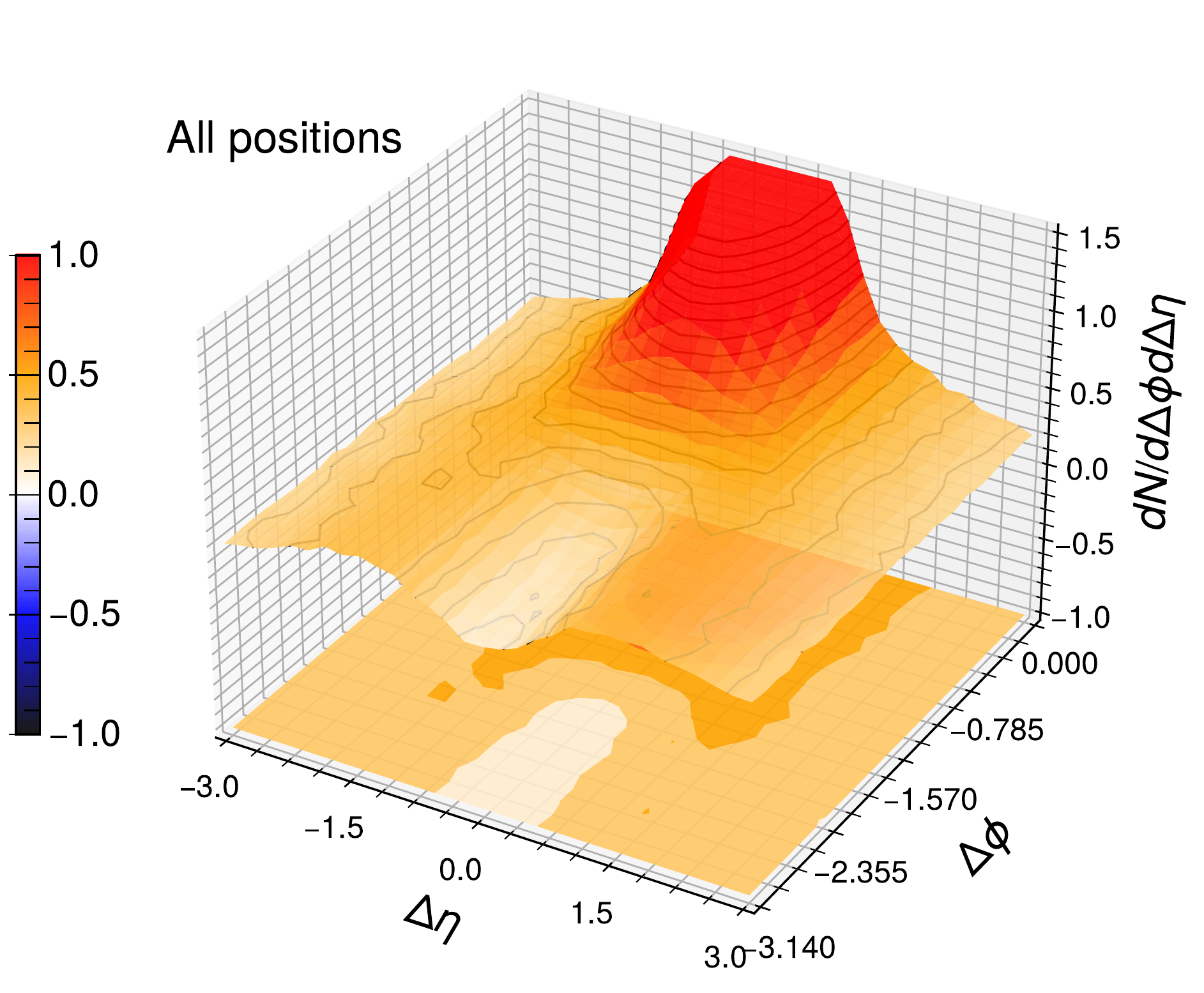}\includegraphics[width=0.45\textwidth]{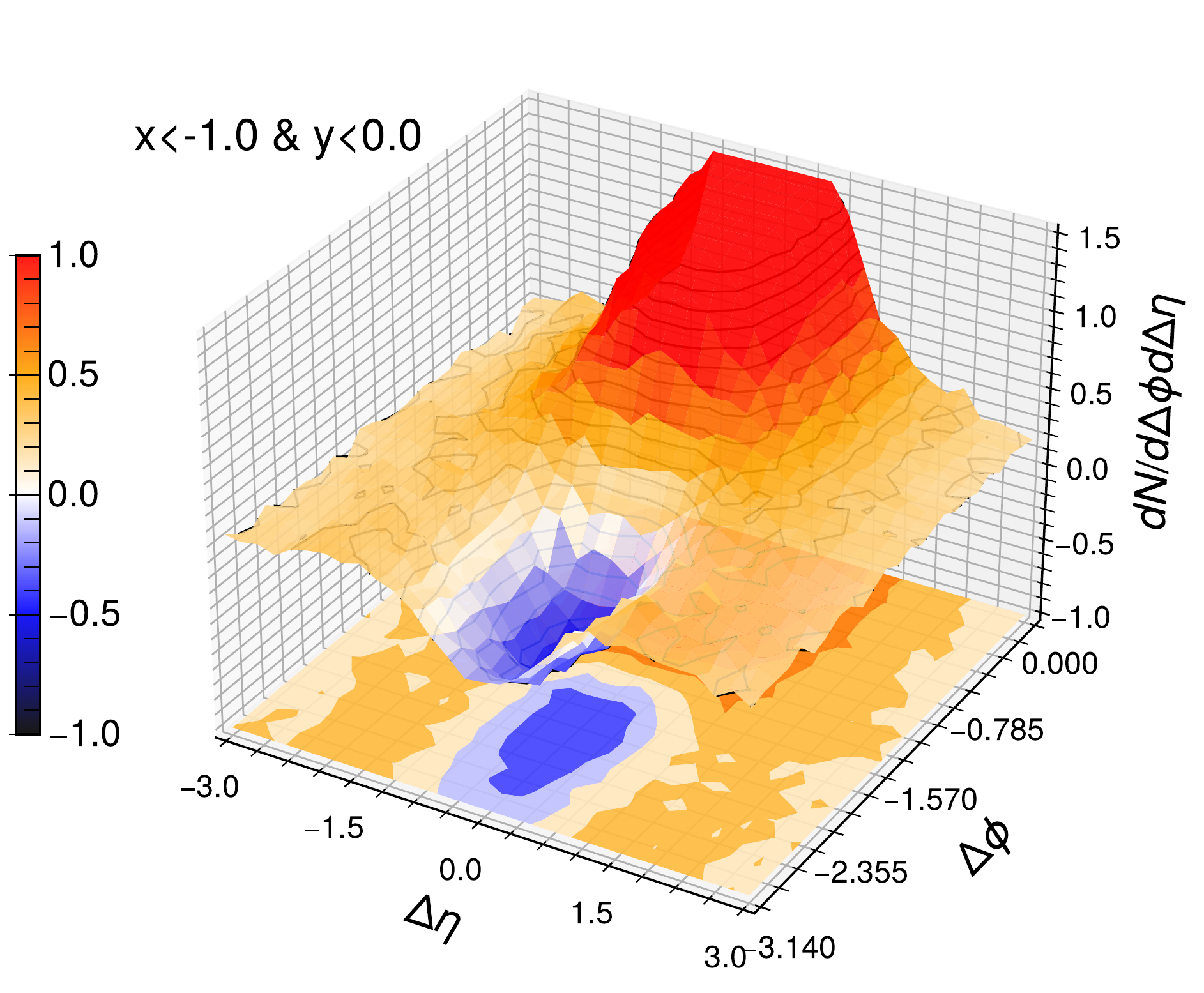}
    \caption{(color online) Distributions in azimuthal angle $\Delta\phi$ and rapidity $\Delta\eta$ of charged particles with $1\le p_T \le 2$ GeV from jet shower and medium response. Events are averaged over all jet production positions (left plot) or restricted to region (x<-1, y<-1) using a deep neural network (right plot).}
    \label{fig:diffusion_wake}
\end{figure}

{\it 3D structure of the diffusion wake--}
Since both jets and the Mach cones are 3-dimensional objects, it is interesting to examine the 3D structure of the signals of Mach cones and the diffusion wake as recently suggested in Ref.~\cite{Yang:2022nei}.  We can do so for $\gamma$-jet events with and without restricted regions of the initial jet production positions. Fig.~\ref{fig:diffusion_wake} shows the 2-dimensional jet-hadron correlations as a function of rapidity $\Delta\eta=\eta_{\rm h}-\eta_{\rm jet}$ and azimuthal angle $\Delta\phi=\phi_{\rm h}-\phi_{\rm jet}$ for soft charged hadrons with $p_T=1-2$ GeV/$c$ from $\gamma$-jets with $p_T^\gamma\in(200,250)$ GeV/$c$, $\left | \eta_\gamma \right |<0.32$ and $p_T^{\rm jet}>100$ GeV/$c$, $\left |\eta_{\rm jet} \right |<0.75$ in 0-10\% central Pb+Pb collisions at $\sqrt{s}=5.02$ TeV. The left plot is for $\gamma$-jet events averaged over all possible initial jet production positions with the jet going in the positive $y$ direction (perpendicular to the reaction plane). There is a peak in the jet direction corresponding to soft hadrons from the Mach cone and the jet shower partons. This peak sits on top of a ridge-like (along the $\Delta\phi$ direction at $\Delta\eta=0$) background from MPI. In the $\gamma$ direction we see a shallow valley on top of the MPI ridge due to the depletion of soft hadrons by the diffusion wake. The minimum point of the diffusion wake valley is in the direction of the trigger $\gamma$ at $\Delta\phi=\pi$ and $\Delta\eta=0$.

We now restrict the initial jet production positions to $(x<-1, y<0)$ with the jet traveling in the positive $y$ direction using the deep neural network predictions. This scenario is very similar to that in the right panel of 
Fig.~\ref{fig:dndphi} (d,h,l) where the jet traverses approximately one half region of the QGP in a direction that is mostly tangential to the radial flow. The structure of the Mach cone and diffusion wake in the 2D jet-hadron correlation function is much more enhanced as shown in the right plot of Fig.~\ref{fig:diffusion_wake}. Restricting the initial jet production positions this way forces the jet to traverse longer path lengths on the average, lose more energy and induce a stronger medium response. The peak in the jet direction is higher as compared to the left plot. The peak position is also shifted away from jet direction (difficult to see here because the tip of the peak is cut-out by the limited range of the vertical axis of the plot) due to the radial flow and density gradient.  The diffusion wake valley in the $\gamma$ direction is much deeper with mostly negative yields \footnote{The jet-hadron correlation is calculated with the normal hydrodynamic background subtracted which is obtained by running the same hydro event without the $\gamma$-jet. Negative yields in the jet-hadron correlation means that the total yield is smaller than the hydro background.}.  Since the jet propagation direction is mostly tangential to the radial flow and the path is off the center of the QGP in this case, the diffusion wake is swiped along the radial flow and drifted by the density gradient away from the $\gamma$ direction. This leads to a diffusion wake valley whose minimum point (the dip) is shifted away from the $\gamma$ direction ($\Delta\phi=\pi$) as we see in the right plot. This 3D structure enhanced by neural network selection of the initial jet production positions is much more dramatic than the plain azimuthal angle dependence of jet-hadron correlations.

\begin{figure}[!htp]
    \centering
    \includegraphics[width=0.6\textwidth]{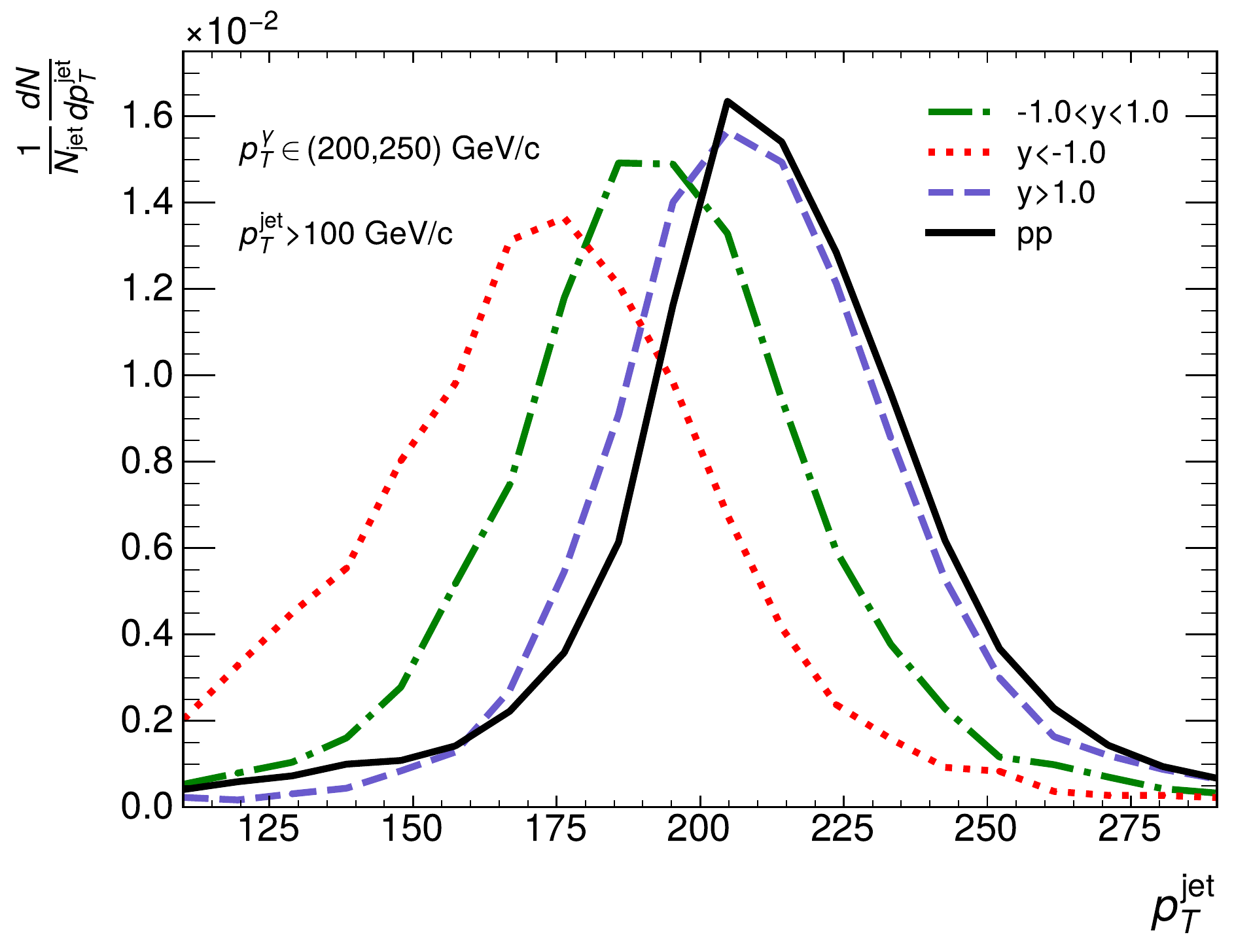}
    \caption{(color online) $p_T$ distribution of $\gamma$-jets in p+p (solid black) and 0-10\% central Pb+Pb collisions at $\sqrt{s}=5.02$ TeV with different path length selected by deep neural network. The initial $\gamma$-jet configurations are generated by PYTHIA8 with $\hat p_{Tmin}=200$ GeV/$c$.}
    \label{fig:gamma_jet_asy}
\end{figure}
{\it The $\gamma$-jet asymmetry--}
Another application of the deep learning assisted jet tomography is to study the length dependence of jet energy loss. Using the predictions of the initial jet production positions by the neural network from the information of jet hadrons, one can classify events with different regions of the initial jet production positions and therefore different averaged jet propagation length. One can then extract the path length dependence of the jet modifications.  Shown in Fig.~\ref{fig:gamma_jet_asy} are jet $p_T$ distributions for $\gamma$-jets that propagate in the positive $y$ direction (perpendicular to the reaction plane) with different path lengths as selected by deep neural network in 0-10\% Pb+Pb collisions at $\sqrt{s}=5.02$ TeV. The solid-black line is the baseline in proton+proton collisions without jet-medium interaction.  The blue-dashed line is for jets with $y>1$ that are produced close to the surface of QGP. They have on the average short path lengths and small jet energy loss. The corresponding jet distribution is slightly modified from the p+p baseline with the peak shifting slightly to smaller $p_T$ indicating a small jet energy loss. If we constrain the predicted jet production positions to $y \in [-1, 1]$ (dash-dotted-green line) and $y < -1$ (dotted-red line),
the average path lengths are increasingly longer with larger jet energy loss. The corresponding jet $p_T$ distributions therefore shift to smaller $p_T$ as shown. These trends of the $\gamma$-jet asymmetry is a clear evidence for the power of the deep learning assisted jet tomography which can be used in the future for more detailed study of jet quenching and jet-induced medium response.

\section{Summary}

In this paper, we have developed a deep learning assisted jet tomography and applied the technique to study jet-induced Mach cones and the associated diffusion wake in high-energy heavy-ion collisions. The deep learning assisted jet tomography is based on a point cloud neural network and allows one to predict the initial jet production locations using information of final state hadrons with $p_T>$ 2 GeV/$c$ inside a jet in each event.   
The neural network is trained with data from CoLBT-hydro model simulations of jet propagation and medium response in high-energy heavy-ion collisions. The network is also tested against pseudo-data from the LIDO model with high accuracy. The mean square root deviation between network prediction and the true jet production position 
is about $2.2\sim 2.4$ fm. This network therefore allows one to select events with approximate regions of the initial jet production locations inside the overlap transverse area of two colliding nuclei with known jet propagation direction relative to the radial flow.

Using the deep learning assisted jet tomography we have investigated the signals of jet-induced Mach cones and the associated diffusion wake in the final jet-hadron correlation for soft hadrons from CoLBT-hydro simulations of $\gamma$-jet events in high-energy heavy-ion collisions. By restricting the initial jet production positions to different regions of the nuclear overlap area, we can enhance the signals of the Mach cone and diffusion wake and investigate the influence of the radial flow and density gradient. 
When the initial jet production positions are symmetric to the jet direction, the jet-hadron azimuthal correlation is shown to get broadened by the wind of the radial flow and drifted diffusion due to the density gradient. The width of the azimuthal correlation and the soft hadron yield are shown to increase with the averaged jet path length when we vary the restricted region of the initial jet production location. For a long jet path length, a clear negative dip due to the diffusion wake is observed behind the conical shock front. If the initial jet production positions are restricted to only one half region of the nuclear overlap area, the final jet-hadron correlation becomes asymmetric with respect to the jet direction due to the distortion of the Mach cone and diffusion wake by the radial flow and density gradient. The Mach-cone peak in the jet-hadron correlation is shifted away from the jet direction and the dip of the diffusion wake is also shifted away from the $\gamma$ direction. Both of these shifts are driven by the blowing wind of the radial flow and drifted diffusion due to the density gradient. We also applied the deep learning assisted jet tomography to look at the signals of the Mach cone and diffusion wake in the 2-dimensional jet-hadron correlation in both rapidity and azimuthal angle, which is more dramatic than in the azimuthal angle jet-hadron correlation. Observing this phenomena in experimental data will provide direct evidence for the existence of Mach cones in heavy ion collisions and their distortion by the radial flow and density gradient.

We also illustrated how one can use the jet tomography to extract path length dependence of jet energy loss from the measured $\gamma$-jet $p_T$ distributions. These are just two examples of how one can use the deep learning assisted jet tomography to study jet quenching and jet-induced medium response in high-energy heavy-ion collisions.

\begin{acknowledgments}

This work was supported in part by the National Science Foundation of China under Grant Nos. 12075098, 11935007, 11221504, 11861131009 and 11890714, by the Director, Office of Energy Research, Office of High Energy and Nuclear Physics, Division of Nuclear Physics, of the U.S. Department of Energy (DOE) under grant No. DE- AC02-05CH11231, by the U.S. National Science Foundation under grant No. OAC- 2004571 within the X-SCAPE Collaboration. This work was also supported by the Laboratory Directed Research and Development (LDRD) Program at Los Alamos National Laboratory, by National Key Research and Development Program of China under Grant No. 2020YFE0202002 and by Fundamental Research Funds for Central Universities in China. Computations are performed at Nuclear Science Computer Center at CCNU (NSC3) and the National Energy Research Scientific Computing Center (NERSC), a U.S. Department of Energy Office of Science User Facility operated under Contract No. DE-AC02- 05CH11231.

\end{acknowledgments}

\bibliographystyle{unsrt}
\bibliography{ref}

\end{document}